\documentclass[referee,psfig,epsfig]{aa}
\usepackage{graphics}
\usepackage{amssymb}
\begin{document}
\thesaurus{ 06() }
\titlerunning{test of Synchrotron Self-Compton process}
\title{VLBI observations of 3C273  at 22\,GHz and 43\,GHz  \\
II: test of Synchrotron Self-Compton process}

\author{F. Mantovani\inst{1}, W. Junor\inst{2}, I.M. M$^c$Hardy\inst{3},
C. Valerio\inst{1}
}
\institute{Istituto di Radioastronomia del CNR, Bologna, Italy
\and Institute for Astrophysics, University of New Mexico, Albuquerque, NM, USA
\and Department of Physics and Astronomy, University of Southampton, UK}
\offprints{fmantovani@ira.bo.cnr.it}
\date{ Submitted  }
\maketitle
\markboth{F. Mantovani et al.}{3C273: Test on SSC}	
\begin{abstract}
The VLBI observations at 22\,GHz and 43\,GHz of the quasar 3C273
obtained during a multi-frequency campaign in late 1992 in the radio,
millimetre and X-ray bands allow us to derive the components' angular
sizes, their peak fluxes and turnover frequencies.  Lower limits to
the Doppler factors have been derived by comparing the observed X-ray 
fluxes with those predicted by the Synchrotron Self-Compton model.
Independent estimates of the Doppler factors were obtained through the
assumption of the energy equipartition between the particles and the
magnetic field.  Of the five components used to model the first two
milli-arcseconds of the jet, apart from the core, two components are
in equipartition and the remaining two, at larger distances from the
core, have large Doppler factors and are mainly responsible for the
X-ray emission due to the Synchrotron Self-Compton process.
\end{abstract}
\section{Introduction}
There is a general agreement that most of the emission from blazars
comes from a relativistic jet oriented close to the observer's
line-of-sight.  In this model, the emission from the radio through to
the UV is synchrotron radiation and it is believed that the X
and $\gamma$-ray emission is due to Compton up-scattering of low
energy seed photons.  The seed photons could arise
externally to the jet, as in the External Compton Scattering or ECS
model, where the photons arise from the accretion disc or broad
line clouds (eg. Melia \& K\"onigl 1989). However the most popular
model is the Synchrotron Self-Compton (SSC) model where the seed
photons are the synchrotron photons generated internally in the
jet, which are then scattered to high energies by their parent
electrons. This model is popular since, at some level, SSC scattering
must occur. 

The SSC and ECS models, and variants thereof, such as the Mirror
Compton Model (eg. Ulrich et al. 1997), can, in principle, be 
distinguished by
careful measurement of the lags, during outbursts, between various
wavebands, principally the high and low energy bands, and by measurement
of the relative amplitude of variability in the high and low energy
bands (eg. Marscher \& Travis 1996; Ghisellini \& Maraschi 1996). However
arranging monitoring in a number of different wavebands is not trivial
and has rarely been achieved (eg M$\rm^{c}$Hardy et al. 1999).

An alternative method of testing the SSC model is
by measurement of the physical parameters of the possible sites of
synchrotron emission by VLBI multi-frequency radio observations of the
cores of compact superluminal radio sources. From such observations we
can measure the angular size, radio spectrum and self-absorption
frequency of the various radio components. If, in addition, the
relativistic Doppler factor has been estimated from the motion of
radio components, then the SSC X-ray flux can be accurately predicted
(e.g.\ Marscher 1983). Thus a combination of VLBI and X-ray observations
can provide an excellent test of the SSC model. Of course, in most
situations some important parameter, usually the Doppler factor,
is not well determined. In that case the Doppler factor can be
very well constrained from the combination of VLBI and X-ray observations.
A consistency check can be achieved by comparing the derived Doppler
factor with its estimate derived by equipartition arguments (Readhead 1994). 
The comparison can enable us to distinguish which, if any, of
the radio components is most likely to be the major source
of X-ray emission.

In this paper, we compare the X-ray and VLBI observations of 3C273,
one of the brightest and best-studied of the superluminal radio
sources, in order to test the SSC model and to determine the
origin of the X-ray emission in 3C273.

The VLBI observations of 3C273 at 22\,GHz and 43\,GHz
presented in Mantovani et al. (1999; hereafter Paper I), were performed 
as part of a multi-frequency campaign carried out from December 12, 1992 
to January 24, 1993. Over this period 3C273 was monitored every 2 days 
with the PSPC of ROSAT in the
$0.1-2.4\,keV$  band (Leach et al. 1995), at nearly 1 day intervals
at 2, 1.1, 0.8, 0.45 mm with the JCMT (Mauna Kea), IRAM and the Kitt
Peak telescope (M$^c$Hardy et al. 1994). No large flares were seen in that
period but variations in both wavebands of $\sim$30\% on few day timescales
are apparent. The X-ray spectrum consists of 2 power-law components with the
harder component dominating above 0.5 keV. There is a very little correlation
between the variability of the soft and hard components. The soft component
does not correlate with the millimeter variations. The hard component may
correlate reasonably with the millimetre variations, leading the millimetre
by about 10 days but the correlation is weak and may be the result of chance
statistical fluctuation. However, M$^c$Hardy et al. (1999) have demonstrated
that a strong relationship does exist between the X-ray emission and 
the infrared band in 3C273, and the infrared is almost certainly just a
tracer for the whole infrared to millimetre synchrotron spectral component.

The jet structure of 3C273 obtained by the VLBI observations, discussed
in Paper I,  does not dramatically change in the observing window of 43 days.
Good evidence was found for
a bending of the jet, beginning inside the first parsec from the core; this
behaviour strongly suggests a precessing jet. The possibility
that the morphology in 3C273 can also be modelled with Kelvin-Helmholtz
instabilities cannot be excluded, however. 

The model-fitting of the jet structure makes it possible to
follow the changes in separation of the components with respect to the
core. Moreover, we can derive the components' angular sizes (Section
2.1) and their peak flux densities and turnover frequencies (Section
2.2). These parameters are fundamental to the analysis in terms of the
SSC process (Section 2.3). In Section 3 the results will be discussed.
\section{Parameters derived from the observations}
\subsection{Angular sizes}
The source 3C273 was observed with an array consisting of the VLBA plus
Effelsberg, Medicina and Noto at 22\,GHz, while the stand-alone VLBA was 
used at 43\,GHz. The structure of the jet of 3C273 has been modelled with
Gaussian components. The model-fitting was performed on the final 
self-calibrated visibilities of each data set (see Paper I). 
The inner two milliarcseconds (mas), have been modelled with five 
components at both 43\,GHz and 22\,GHz with a good formal agreement, 
$\chi_{\nu}^2$,  between data and model.
The ratio between the total flux density from the hybrid images and the
flux density obtained adding up the flux density of each individual
components in the related model was always close to 1. 
The models obtained from the 22\,GHz data sets, using the visibilities  
in the baseline range $\leq$450 M$\lambda$ to match the resolution to 
that of the 43\,GHz observations, will also be used in the following analysis.

The Gaussian components found in the models are compact compared to
the synthesized beam and their angular sizes do not change noticeably between 
epochs.  Consequently, the mean value of the angular sizes found for each 
component will be adopted.
The major axis and the minor axis FWHMs, $\theta_1$
and $\theta_2$ respectively, were then averaged geometrically.  
To be conservative in the calculation of the X-ray emission by the SSC 
process, the mean angular sizes of the components are obtained with 
the relation
$\theta=1.8\sqrt{\theta_1\theta_2}$, to convert the deconvolved Gaussian
diameters to the optically thin sphere diameter.  The angular sizes
in Table 1 computed for each component of the 22\,GHz and the 43\,GHz
models respectively, are similar.  In order to compute the X-ray emission
by SSC process, in the following we will adopt the
angular sizes obtained by modelfitting the 43\,GHz data. The observations
at that frequency had a better {\it uv} coverage and comparable
resolution in the North-South and East-West directions; this allows us to
derive more reliable parameters. The mean flux densities values for each
of the components is also reported in Table\,1. 

The Table 1 is organized as follows: column 1, component's label; 
column 2 and 3, minor and major axis FWHMs at 22\,GHz respectively;
column 4, angular sizes at 22\,GHz; 
column 5 and 6, minor and major axis FWHMs at 43\,GHz respectively; 
column 7, angular sizes at 43\,GHz; 
column 8 and 9 mean separation and Position Angle of each component
 from component 1 taken as the reference point;
column 10 and 11, mean flux density of the components at 22\,GHz and 43\,GHz 
 respectively (see below);
column 12, spectral index $\alpha$ ($S\propto \nu^{-\alpha}$) between the two
 frequencies.
\begin{table}[h]
\centerline{\bf Tab. 1 - Component parameters}
\vspace{0.5cm}
\hspace{0.5cm} 
\begin{tabular}{cccccccrrllr}
\hline
comp & $<\theta_1>$&$<\theta_{2}>$ & $<1.8\sqrt{\theta_1\theta_2}>$&$<\theta_1>$&$<\theta_{2}>$&$<1.8\sqrt{\theta_1\theta_2}>$ & $<r>$ & $<PA> $ &
S$_{22}$ & S$_{43}$      & $\alpha$      \\
  & (mas) & (mas) & 22 GHz (mas)  & (mas) & (mas) & 43 GHz (mas) & (mas) & (deg) & (Jy) & (Jy)      &   \\
\hline
  1  & 0.15 & 0.30 & 0.38 & 0.16& 0.34& 0.42 &      &      & 1.82 & 2.57 & --0.52 \\
  2  & 0.19 & 0.41 & 0.50 & 0.15& 0.33& 0.40 & 0.34 & --100& 4.50 & 2.02 &   1.20 \\
  3  & 0.23 & 0.45 & 0.58 & 0.29& 0.48& 0.67 & 0.70 & --115& 8.46 & 4.45 &   0.97 \\
  4  & 0.28 & 0.44 & 0.63 & 0.25& 0.49& 0.63 & 1.29 & --121& 2.76 & 0.83 &   1.81 \\
  5  & 0.27 & 0.47 & 0.64 & 0.25& 0.41& 0.57 & 1.75 & --108& 3.05 & 1.06 &   1.60 \\
\hline
\end{tabular}
\vspace{0.5cm}
\end{table}
\subsection{Radio spectrum}
The X-ray emission due to SSC is strongly dependent on the 
self-absorption turnover frequency 
$\nu_m$ and on the flux density S$_m$ at $\nu_m$.
The present VLBI observations at 22\,GHz and 43\,GHz
provide simultaneous flux densities measurements with similar resolving power.
Simultaneous observations with similar resolution at
lower frequency (i.e. VSOP observations at 5\,GHz), as are possible now,
would have been ideal for a proper estimate of both $\nu_m$ and S$_m$. These 
parameters will be derived here with the following approach.

The total power flux densities measurements of 3C273 at various
frequencies made in the period of our observations were taken from the
compilation by von Montigny et al.\ (1997). The appropriate flux
density measurements for the arcsecond scale jet, as given by Conway
et al. (1993), have been subtracted from the total flux measurements
since the arcsecond jet has an optically-thin spectrum and its flux density
should be fairly constant in time. What
remains is the emission from the core and milliarcsecond jet at the
time of our observations. The spectra are plotted in Fig.\,1 and 2.
\begin{figure*}
\resizebox{12cm}{!}{\includegraphics{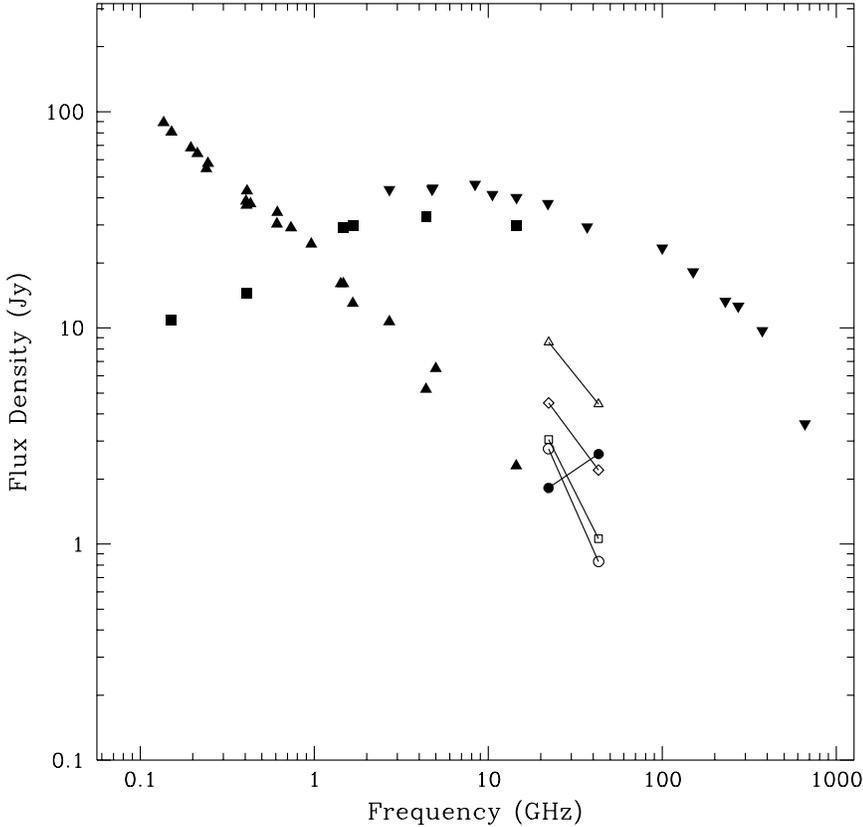}}
\hfill
\caption[]{ Radio spectrum of 3C273. Symbols: $\blacktriangle$ flux density 
from the arcsecond scale jet as in Conway et al. (1993), $\blacktriangledown $ 
total flux density as from von Montigny et al. (1997), 
$\blacksquare$ flux density from
the arcsec core.  The flux densities from the various components are
$\bullet$ comp. 1, $\diamondsuit$ comp. 2, $\triangle$ comp. 3,
$\circ$ comp. 4, $\sq$ comp. 5 respectively.
}
\end{figure*}
%
%
\begin{figure*}
\resizebox{12cm}{!}{\includegraphics{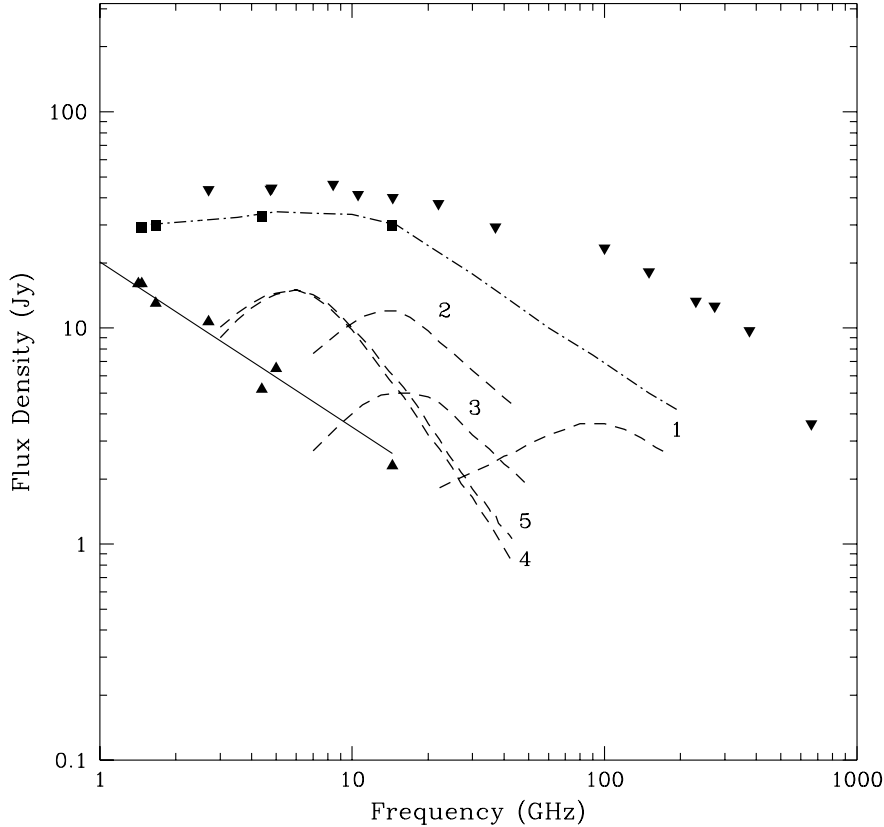}}
\hfill
\caption[]{ Decomposition of the spectrum of 3C273. 
 Symbols: $\blacktriangle$ flux density from
the arcsecond scale jet as in Conway et al. (1993), 
$\blacktriangledown $ total flux density as from von Montigny et al. (1997), 
$\blacksquare$ flux density from the arcsec core.
The spectra of the five components are plotted.
The dotted line represents the spectra sum of the five components.
}
\end{figure*}
The integrated spectrum of the core and milliarcsecond jet
has been reconstructed, by adding up the spectra which the five 
components should have, given the VLBI flux density measurements 
which we obtained at the two frequencies.
The self-absorption turnover frequency $\nu_m$ and flux density S$_m$ from
the spectral decomposition are reported in Table 2.

Component 1 is the only one which shows an inverted spectra and should be the 
core. Its turnover frequency of $\sim$90 GHz is confirmed by quasi-contemporary
VLBI observations at 86\,GHz made on January 1993 (T. Krichbaum, private 
communication) in which the unresolved core component has a flux density 
of 3.6 Jy. The 86\,GHz model shows two more components at a separation of
0.39 mas and 1.4 mas from the core respectively. Their positions correspond 
to the positions of components
3 and 4 in our model(s) and they have flux densities of 2.8 Jy and 0.9 Jy
respectively. These last values fit with the measured trend for the 
optically-thin part of the spectrum of those components. Component 2 is 
weaker than component 3 and the turnover frequencies of both components
are at $\sim$15\,GHz.
Components 4 and 5 have spectra with turnovers at much lower 
frequencies than components 2 and 3 and the optically-thin 
spectra have steep gradients.
The total correlated flux density at both 22\,GHz and 43\,GHz does not 
account for the expected emission at arcsecond scale resolution from the 
core at those frequencies. The missing flux density must come from the region 
of the jet intermediate between the mas and the arcsecond scales;
this part cannot be properly imaged by the available interferometers.
A full discussion of the amplitude calibration can be found in Paper\,I.
\subsection{X-ray emission by Synchrotron Self-Compton process}
Monitoring observations from radio through millimeter wavelengths have shown
many blazars have a characteristic behaviour -- flares with a delayed
broader peak relative to the higher frequencies, with a slow decline 
thereafter.
From X-ray monitoring observations on 3C345, Unwin et al.\ (1997) found that
the decline in X-ray flux is correlated with the radio flux at frequencies
above 37\,GHz. This correlation is strong evidence that X-ray emission
originates in the parsec scale jet. In Paper I, we showed that  
the parsec-long radio jet in 3C273 could be modelled with five 
well-separated components. 
Let us assume that these components are homogeneous and spherical,
with tangled magnetic fields and high-energy electrons which follow
an isotropic energy distribution ($N(E)=N_0 E^{-(2\alpha+1)}$). The electrons
produce radio emission by synchrotron radiation and produce at least part of
the X-ray flux by inverse Compton scattering of their own synchrotron  
photons. The electron in the jet may also produce X-rays by inverse Compton
scattering of other low energy seed photons, eg from nuclear accretion disc,
or by scattering synchrotron jet photons which have been reflected back
to the jet from nearby gas clouds (Ulrich et al. 1997).

The monochromatic SSC X-ray flux at a frequency $\nu_X$ (keV) produced by 
such a synchrotron source with angular diameter $\theta$ (mas), 
redshift {\it z}, spectral index $\alpha$, break frequency $\nu_2$ (GHz),
self-absorption turnover frequency $\nu_m$ (GHz) and flux density
$S_m$ (Jy) is (see Marscher 1983):
\begin{equation}
S_X = h(\alpha) ln(\frac{\nu_2}{\nu_m}) \theta^{-2(2\alpha+3)} 
\nu_m^{-(3\alpha+5)}
 S_m^{2(\alpha+2)} \nu_X^{-\alpha} [\frac{1+z}{\delta}]^{2(\alpha+2)} \mu Jy
\end{equation}
where $h(\alpha)$ is a dimensionless parameter depending on the 
spectral index ($\log{h(\alpha)} \approx -0.62 \alpha + 2.24$, with 
$0.4 \leq \alpha \leq 1.0$) and $\delta$ is the Doppler factor.
A value of 1000\,GHz was adopted for $\nu_2$ (see von Montigny et al 1997). 

The resulting expected monochromatic X-ray fluxes (in $\mu$Jy) from
each component are given in Table\,2, and have been computed assuming
component spectral indices equal to 1 and Doppler factors equal to 1.
In Table\,2 we also give the expected fluxes, in erg/(sec cm$^2$), 
integrated over the 0.1--2.4 keV ROSAT observing band, i.e.
\begin{equation}
S_X(0.1-2.4~KeV)= f(\alpha) S_{1keV} (\mu Jy) \times 10^{-29} ergs/(cm^2 sec)
\end{equation}
where f$(\alpha)$ is a function of the spectral index which becomes 
7.6$\times 10^{17}$ for $\alpha=1$, with the assumption that
$<\nu> = 2.4 \times 10^{17}$ Hz, i.e. an energy of 1 keV.
\subsection{Doppler factors from SSC}

The predicted Inverse Compton X-radiation should not exceed the observed 
X-ray flux. Since we have simultaneous radio and X-ray measurements,
lower limits to the Doppler factors for each of the five components  
in the parsec-scale jet can be derived.
The self-Compton X-ray fluxes for components 1 to 5 have been calculated
and have been compared to the X-ray flux detected by ROSAT.
Leach et al. (1995) have noted 
a mean variability on  the counts rate of $\sim$8\% on short, two day
time scales. Such a fluctuation is ignored in the following discussion.
Those authors have also pointed out the lack of correlation between the 
variations of the
X-ray emission in the 0.1--0.3\,keV and 1.5--2.4\,keV, which might indicate
that two distinct mechanisms of emission exist. They suggest that 
the X-ray spectrum is formed by two components with different power law 
spectra with
the same intensity at about 1\,keV. The soft component is
probably due to inverse Compton scattering of UV photons from the accretion
disk, while the hard component with spectrum $\sim$0.5 is due to  
inverse Compton scattering of their own
synchrotron radiation by the electrons in the relativistic jet.
This hypothesis is supported by two facts: (1) the
mm-band emission during the multi-band campaign (M$^c$Hardy et al. 1994) 
is possibly correlated to the light curve of the `hard' 
component of the X-ray emission; this component is probably due to SSC
and (2) correlated IR/X-ray variability in 3C273
(M$^c$Hardy et al.\ 1999). Thus 
we consider only the hard component ($S_{Xhard}=7.7\times10^{-11} erg/cm^2s$)
to be due to the SSC process 
in order to obtain lower limits to the Doppler factors listed in
Tab.\,2. 

The ratio between the X-ray flux density expected from SSC 
($S_X(0.1-2.4 keV$) and the X-ray emission detected by ROSAT, 
is proportional to
$(\delta_X)^{2(\alpha+2)}$, where $\delta_X$ is the Doppler factor.
The spectral index for each component is listed in Table\,1; a spectral 
index $\alpha=0$ was adopted for component 1 (the core).
Furthermore, the calculated X-ray flux
cannot be larger than the observed value, so equating the calculated and
measured flux densities yields $\delta_{min}$, a lower limit to
$\delta$. The limits obtained for $\delta_X$ are reported in Table\,2.
\subsection{Doppler factors from equipartition}
An independent method  of estimating  the Doppler factor in powerful
extragalactic radio sources is the use of the equipartition condition.
Since we know the components' angular diameters, turnover frequencies and flux 
densities, it is possible to compute the energy
density content in the relativistic electrons {\it u$_e$} and in the 
magnetic field {\it u$_m$}. If we assume that the source is in equipartition
(i.e. Readhead 1994), the Doppler factors of the components can be derived 
from the relation:  
\begin{equation}
 E_{em} = \frac{u_e}{u_m} \simeq 3.0 \times 10^{10} (\frac{H_{\circ}}{100})
          D^{-1} \nu_m^{-17.5+\alpha} (GHz) \theta^{-17} (mas)
          S_m^{8} (Jy) \frac{(1+z)^9}{\delta^7}
\end{equation}
where, of course, $ E_{em}=1$ and in which:
\begin{equation}
 D = \frac{q_{\circ}z + (q_{\circ} - 1)(\sqrt{1+2q_{\circ}z}-1)}
             {q_{\circ}^2}
\end{equation}
where ($H_{0}=100\,km\,s^{-1}\,Mpc^{-1}$, $q_{0}=0.5$).
The Doppler factors from equipartition $\delta_{eq}$ are reported in Table 2.
\subsection{Brightness Temperatures of the components}
The lower limits to the brightness temperature, $T_b$, for each of the 
components
along the jet of 3C273 can be computed with the expression:
\begin{equation}
 T_{b} = 1.41 \times 10^9 (1+z) (\frac{S_m}{Jy}) 
       (\frac{\theta}{mas^2})^{-2} (\frac{\lambda_m}{cm})^2
\end{equation}
where $\theta=1.8\sqrt{\theta_1\theta_2}$ is the components' angular size
and $\theta_1$ and $\theta_2$ are the mean minor and the mean major axis 
angular sizes of each component,
$S_m$ is the turnover flux density, $\lambda_m$ the turnover wavelength 
and $z$ is the redshift. 
The derived values are listed in Table 2, together with the 
brightness temperature from equipartition $\left ( T_{b} \right )_{eq}$
obtained by the equation (4$b$) in Readhead (1994) when the Doppler factor
of the emission region is taken to be equal to 1.

As expected, the equipartition requirements give a limit to the
brightness temperature of $\simeq 10^{11}\,^{\circ}$K, while the values
for the brightness temperature in the rest frame of the source $T_b$,
given by equation (5), are in the range from $\sim 10^{10}\,^{\circ}$K for
the core to values slightly exceeding the IC limits for the outer 
components 4 and 5.
\begin{table}[h]
\centerline{\bf Tab. 2 - Parameters and derived physical quantities for each
of the components}
\vspace{0.5cm}
\hspace{0.5cm} 
\begin{tabular}{cccllllcc}
\hline
comp & $\nu_m$ & S$_m$  &S$_X$(1keV)     & S$_X$(0.1-2.4keV) 
     & $\delta_X$  & $\delta_{eq}$ &$T_{b}$&$\left ( T_{b} \right )_{eq}$   \\
     &  GHz    &  Jy    & $\mu$Jy        & erg/(sec cm$^2$)        &               &                 & $^\circ$K & $^\circ$K   \\
\hline
  1  &  90     &  3.5   & 4.5$\times 10^{-7}$& 3.3$\times 10^{-18}$& $>0.01$          & 0.01                 & 3.0$\times 10^{9}$& 4.3$\times 10^{10}$ \\
  2  &  15     &  5     & 16            & 1.2$\times10^{-10}$     & $>1.1$        & 2.2               & 1.8$\times 10^{11}$ & 4.1$\times 10^{10}$ \\
  3  &  14     & 12     & 35            & 2.7$\times10^{-10}$     & $>1.2$         & 2.5               & 1.7$\times 10^{11}$ & 5.2$\times 10^{10}$ \\
  4  &   6     & 15     & 2.5$\times10^5$& 1.9$\times10^{-6}$      & $>3.8$          & 29.5              & 1.3$\times 10^{12}$ & 5.5$\times 10^{10}$ \\
  5  &   6     & 15     & 2.2$\times10^5$& 1.6$\times10^{-6}$      & $>4.0$         & 37.4              & 1.6$\times 10^{12}$ & 5.5$\times 10^{10}$ \\
\hline
\end{tabular}
\vspace{0.5cm}
\end{table}
\subsection{Apparent speed, $\gamma$ factor and the angle to the line of 
sight}
From the observations presented in Paper I, we were able to measure the
apparent speeds of the inner jet components in orthogonal directions. The
values obtained, reported in Table\,3, are generally consistent with those
available in literature (i.e. Zensus et al. 1990).
The apparent speed $\beta_{app}= v_{app}/c$ and the
apparent $\gamma$ factor can be
calculated. They are listed in Table\,3. That calculation is, of course,
valid only if the apparent component velocities are due to
bulk motion of the components and not if the motions
represent enhanced emission regions caused by the movement of
Kelvin-Helmholtz instabilities along the jet.
\begin{table}[h]
\centerline{\bf Tab. 3 - Derived quantities}
\vspace{0.5cm}
\hspace{0.5cm} 
\begin{tabular}{ccrrr}
\hline
comp & $\vert{\it r}\vert$ & $\beta_{app}$ & $\beta_{min}$ & $\gamma_{min}$ \\
     & mas/yr              & mas/yr      &               &                \\
\hline
  2  &  0.82               & 5.5           & 0.984         &  5.6  \\
  3  &  0.44               & 3.0           & 0.949         & 3.2 \\
  4  &  1.57               & 10.6          & 0.996         & 10.6 \\
  5  &  1.51               & 10.2          & 0.995         & 10.2 \\
\hline
\end{tabular}
\vspace{0.5cm}
\end{table}
Using the lower values for $\beta_{app}$ and $\gamma_{min}$ gives the
angle of the jet respect to the line of sight $\theta_{obs}$ in the range
$15^\circ\leq\theta_{obs}<25^\circ$. Using the higher values we obtain
$2^\circ\leq\theta_{obs}<10^\circ$. 

From these values it is possible to estimate the Doppler factor range
with the expression
\begin{equation}
 \delta = [\gamma_{min}(1 - \beta_{min} cos\theta)]^{-1}
\end{equation}
With $\beta_{min}=0.949$ and $\theta_{obs}=25^\circ$ we obtain $\delta=2.2$ and
with $\beta_{min}=0.996$ and $\theta_{obs}=2^\circ$ we obtain $\delta=20.5$.
\section{Discussion}
M$^c$Hardy et al.\ (1999) have recently shown that the medium energy X-ray
(3--20\,keV) from the NASA Rossi X-ray Timing Explorer observations and
the near infrared fluxes from the UK Infrared Telescope (UKIRT) in the quasar
3C273 are higly correlated. The lag between the IR and the X-ray bands is very
small but, in at least the first flare observed by M$^c$Hardy et al., 
the IR leads the X-ray emission by
$\sim 0.75\pm0.25$ days. This lag rules out the External Compton process but 
is consistent with the SSC model or, possibly, the Mirror Compton model. 

On the assumption that the X-ray emission is produced by the SSC mechanism
we have combined the radio observations with the contemporary X-ray
observations to derive physical parameters of the milliarcsecond scale jet
in 3C273, particularly the Doppler factors of the radio components.

There is other evidence that the X-ray emission in blazars originates in a 
parsec scale jet. For example, in the case of 3C345, Unwin et al.\ (1997) 
show that the X-ray flux density measurements are directly correlated with 
the total flux density at frequencies above 37\,GHz, both in amplitude range 
and variability. The predicted X-ray emission strongly depends on the
angular size, turnover frequency and flux density of the radio components.
Of those parameters, the angular size is the only one which comes directly
from the present observations. It is worth noting that the usefulness of
those measurements are affected by practical limitations.
The maximum available baseline length, for example, limits the 
achievable resolution  of the interferometer, so that the given angular sizes 
can be considered as upper limits only. 

The turnover frequency and flux density for each of the components are more
difficult to determine. The procedure followed in section 2.2 allows us
to estimate the arcsecond core spectrum of 3C273 at the time of these
VLBI observations using the mas components along the jet. The amplitude 
calibration of these observations has been discussed in Paper I. The missing 
flux at high frequencies should come from the extended structure of the jet, 
resolved out by these VLBI observations. 
The emission from the five components conspires in such a way that
the almost flat spectrum of the core is reproduced.

The most relevant results from the SSC model and equipartition 
arguments, which are listed in Table\,2, allow us to say that: 

(a) component 1 (the core) is probably inhomogenous and there are no 
indications of relativistic effects. The X-ray emission from the core
is rather weak but the computation has been done considering the core
as an homogeneous expanding sphere. However, Unwin et al.\ (1985, 1997)
have shown that the  cores in 3C273 and in 3C345 cannot be the origin of
their X-ray emissions, even applying the probably more relevant
inhomegeneous-jet model of K\"onigl (1981). The model under-estimates the
X-ray flux from the core for any possible combination of parameters 
derived from the observations.

(b) components 2 and 3 show Doppler factors from SSC and equipartition that are
similar and close to 1. This could mean that components 2 and 3
are stationary, are the origin of the observed X-ray emission, and are close
to equipartition. Alternatively if the Doppler factor for these components is
$>1$, as seems reasonable given that the derived Doppler factors for all of
the other components apart from the core are $>1$, then these components
cannot be important sources of X-ray emission.

(c) components 4 and 5 show very large valus of $\delta_X$. It follows that
   relativistic effects must be present. Moreover, $\delta_X$ and 
   $\delta_{eq}$ are rather different from each other, indicating that the 
   components might not be in equipartition.

(d) the brightness temperatures for components 4 and 5 slightly exceed 
the Compton limits. These results and those in (c) are mainly due to the 
low values for $\nu_m$ that result
from the decomposition of the spectrum. It might be noted that these
two components do have rather steep optically-thin spectral indices.
As mentioned in chapter 2.2, the steepening of the spectrum of component 4
is supported by observations at 86\,GHz. Moreover, a less steep spectrum
implies a shift of the turnover frequency to even lower frequencies,
to account for the arcsecond flux density.

The X-ray emission probably comes mainly from components 4 and 5. Those two 
components are at the largest separation from the core.  Similar results 
are found by Unwin (1997) for 3C345, in which component C7, and not the
unresolved core region, is the dominant source of
X-ray emission in the period of their observations. 

The present results are in a similar vein. The X-ray emission increases with
the separation of the components from the inhomogeneous core. With 
increasing separation from the core region, the components show a decreasing
value for the turnover frequency and an increase in the turnover flux density.
The more distant components have much higher values for the Doppler factor
derived from equipartition arguments than derived from SSC modelling and so
are very likely not in equipartition.
The increase in values of both SSC and equipartition
Doppler factors might simply mean that the jet axis bends closer towards
the observers line of sight at larger distances from the core.
\acknowledgements
The authors would like to thank Roberto Fanti for many useful discussions
and Thomas Krichbaum for the 86\,GHz VLBI model for 3C273 kindly provided
prior to publication.
%
%
%

%
\end{document}